%% file: costpltl.tex
\documentclass[submission,copyright,creativecommons, fleqn]{eptcs}
\providecommand{\event}{GandALF 2015} 

\input{preamble}

\title{Parameterized Linear Temporal Logics Meet Costs:\\ Still not Costlier than LTL\thanks{Supported by the project “TriCS” (ZI 1516/1-1) of the German Research Foundation (DFG).}}

\author{Martin Zimmermann
\institute{Reactive Systems Group, Saarland University, 66123 Saarbrücken, Germany}
\email{zimmermann@react.uni-saarland.de}
}

\def\titlerunning{Parameterized Linear Temporal Logics Meet Costs: Still not Costlier than LTL}

\def\authorrunning{M.~Zimmermann}

\begin{document}
\maketitle

\begin{abstract}
\input{abstract}
\end{abstract}

\section{Introduction}
\label{sec_intro}

\input{intro}

\section{Parametric LTL with Costs}
\label{sec_defs}
\input{defs}

\section{The Alternating-Color Technique for Costs}
\label{sec_altcolor}
\input{altcolor}

\section{Model Checking}
\label{sec_mc}
\input{mc}

\section{Infinite Games}
\label{sec_games}
\input{games}

\section{Parametric LDL with Costs}
\label{sec_ldl}
\input{cpldl}

\section{Multiple Cost Functions}
\label{sec_mult}
\input{multcost}

\section{Optimization Problems}
\label{sec_opt}
\input{optimization}

\section{Conclusion}
\label{sec_conc}
\input{conc}

\bibliographystyle{eptcs}
\bibliography{biblio}

%

\end{document}

%% file: preamble.tex
\usepackage{amsmath}
\usepackage{amssymb}
\usepackage{amsthm}
\usepackage[utf8]{inputenc}
\usepackage{tikz}
\usetikzlibrary{automata}
\usepackage{todonotes}

\newcommand{\myquot}[1]{``#1''}

\newtheorem{lemma}{Lemma}
\newtheorem{definition}{Definition}
\newtheorem{theorem}{Theorem}

\newtheorem{corollary}{Corollary}
\newtheorem{example}{Example}

\newcommand{\coloneq}{\mathop{:=}}

\newcommand{\cceq}{\mathop{::=}}

\renewcommand{\epsilon}{\varepsilon}

\newcommand{\pow}[1]{2^{#1}}
\newcommand{\nats}{\mathbb{N}}
\newcommand{\card}[1]{|#1|}
\newcommand{\size}[1]{\card{#1}}
\newcommand{\set}[1]{\{#1\}}

\newcommand{\aut}{\mathfrak{A}}

\newcommand{\arena}{\mathcal{A}}
\newcommand{\game}{\mathcal{G}}

\newcommand{\halfthinspace}{{\kern .08333em}}

\newcommand{\ttrue}{\texttt{tt}}
\newcommand{\ffalse}{\texttt{ff}}
\newcommand{\F}{\mathbf{F}}
\newcommand{\G}{\mathbf{G}}

\newcommand{\U}{\mathbf{U}}
\newcommand{\X}{\mathbf{X}}
\newcommand{\R}{\mathbf{R}}

\newcommand{\conc}{\,;}
\newcommand{\Var}{\mathcal{V}}
\newcommand{\cl}{\mathrm{cl}}
\newcommand{\var}{\mathrm{var}}
\newcommand{\varG}{\mathrm{var}_{\mathbf{G}}}
\newcommand{\varF}{\mathrm{var}_{\mathbf{F}}}

\newcommand{\ddiamond}[1]{\langle #1 \rangle\,}
\newcommand{\bbox}[1]{[\halfthinspace #1 \halfthinspace]\,}

\newcommand{\ddiamondle}[2]{{\langle\/ #1 \/\rangle}_{\le #2}\,}
\newcommand{\bboxle}[2]{{[\halfthinspace#1\halfthinspace]}_{\le #2}\,}

\newcommand{\Rexp}{\mathcal{R}}
\newcommand{\rel}[1]{\mathrm{rel}(#1)}

\newcommand{\ltl}{\text{LTL}}

\newcommand{\pltl}{\text{PLTL}}
\newcommand{\pltlcf}{\pltlc_{{\mathbf{F}}}}
\newcommand{\pltlcg}{\pltlc_{{\mathbf{G}}}}
\newcommand{\pltlc}{\text{cPLTL}}

\newcommand{\prompt}{\text{PROMPT}$\textendash$\ltl}

\newcommand{\pldl}{\text{PLDL}}
\newcommand{\pldlc}{\text{cPLDL}}
\newcommand{\ldl}{\text{LDL}}

\newcommand{\pspace}{\textsc{{PSpace}}}
\newcommand{\twoexp}{\textsc{{2ExpTime}}}

\newcommand{\trace}{\mathrm{tr}}

\newcommand{\sys}{\mathcal{S}}

\newcommand{\mem}{\mathcal{M}}

\newcommand{\update}{\mathrm{upd}}
\newcommand{\nextmove}{\mathrm{nxt}}

\newcommand{\cst}{\text{cst}}
\newcommand{\inc}{\kappa}

\newcommand{\mpltlc}{\text{mult-}\pltlc}
\newcommand{\mpldlc}{\text{mult-}\pldlc}

%% file: abstract.tex
We continue the investigation of parameterized extensions of Linear Temporal Logic (LTL) that retain the attractive algorithmic properties of LTL: a polynomial space model checking algorithm and a doubly-exponential time algorithm for solving games. Alur et al.\ and Kupferman et al.\ showed that this is the case for Parametric LTL (PLTL) and PROMPT-LTL respectively, which have temporal operators equipped with variables that bound their scope in time. Later, this was also shown to be true for Parametric LDL (PLDL), which extends PLTL to be able to express all $\omega$-regular properties. 

Here, we generalize PLTL to systems with costs, i.e., we do not bound the scope of operators in time, but bound the scope in terms of the cost accumulated during time. Again, we show that model checking and solving games for specifications in  PLTL with costs is not harder than the corresponding problems for LTL. Finally, we discuss PLDL with costs and extensions to multiple cost functions. 

%% file: intro.tex
Parameterized linear temporal logics address a serious shortcoming of Linear-temporal Logic ($\ltl$): $\ltl$ is not able to express timing constraints, e.g., while $\G(q \rightarrow \F p)$ expresses that every request~$q$ is eventually answered by a response~$p$, the waiting time between requests and responses might diverge. This is typically not the desired behavior, but cannot be ruled out by $\ltl$.

To overcome this shortcoming, Alur et al.\ introduced parameterized $\ltl$~\cite{AlurEtessamiLaTorrePeled01}, which extends $\ltl$ with parameterized operators of the form $\F_{\le x}$ and $\G_{\le y}$, where $x$ and $y$ are variables. The formula~$\G(q \rightarrow \F_{\le x} p)$ expresses  that every request is answered within an arbitrary, but fixed number of steps~$\alpha(x)$. Here, $\alpha$ is a variable valuation, a mapping of variables to natural numbers. Typically, one is interested in whether a $\pltl$ formula is satisfied with respect to some variable valuation. For example, the model checking problem asks whether a given transition system satisfies a given $\pltl$ specification~$\varphi$ with respect to some $\alpha$, i.e., whether every path satisfies $\varphi$ with respect to $\alpha$. Similarly, solving infinite games amounts to determining whether there is an $\alpha$ such that Player~$0$ has a strategy such that every play that is consistent with the strategy satisfies the winning condition with respect to $\alpha$. Alur et al.\ showed that the $\pltl$ model checking problem is $\pspace$-complete. Kupferman et al.\ later considered $\prompt$~\cite{KupfermanPitermanVardi09}, which can be seen as the fragment of $\pltl$ without the parameterized always operator, and showed that $\prompt$ model checking is still $\pspace$-complete and that $\prompt$ realizability, an abstract notion of infinite game, is $\twoexp$-complete. While the results of Alur et al. relied on involved pumping arguments, the results of Kupferman et al.\ where all based on the so-called alternating-color technique, which basically allows to reduce $\prompt$ to $\ltl$. Furthermore, the result on realizability was extended to infinite games on graphs~\cite{Zimmermann13}, again using the alternating-color technique.

Another serious shortcoming of $\ltl$ (and its parameterized variants) is their expressiveness: $\ltl$ is equi-expressive to first-order logic with order and thus not as expressive as $\omega$-regular expressions. This shortcoming was addressed by a long line of temporal logics~\cite{GiacomoVardi13, LeuckerSanchez07, Vardi11, VardiWolper94, Wolper1983} with regular expressions, finite automata, or grammar operators to obtain the full expressivity of the $\omega$-regular languages. One of these logics is Linear Dynamic Logic ($\ldl$), which has temporal operators~$\ddiamond{r}$ and $\bbox{r}$, where $r$ is a regular expression. For example, the formula~$\bbox{r_0}(q \rightarrow \ddiamond{r_1} p)$ holds in a word~$w$, if for every request at a position~$n$ such that $w_0 \cdots w_n$ matches $r_0$, there is a position~$n' \ge n$ such that $p$ holds at $n'$ and $w_n \cdots w_{n'}$ matches $r_1$. Intuitively, the diamond operator corresponds to the eventuality of $\ltl$, but is guarded by a regular expression. Dually, the box-operator is a guarded always. 
Although $\ldl$ is more expressive than $\ltl$, its algorithmic properties are similar: model checking is $\pspace$-complete and solving games is $\twoexp$-complete~\cite{Vardi11}. 

All these logics tackle one shortcoming, but not both simultaneously. This was achieved for the first time by adding parameterized operators to $\ldl$. The logic, called parameterized $\ldl$ ($\pldl$)~\cite{FaymonvilleZimmermann14,FaymonvilleZimmermann15}, has additional operators $\ddiamondle{r}{x}$ and $\bboxle{r}{y}$ with the expected semantics: the variables bound the scope of the operator. And even for this logic, which has parameters and is more expressive than $\ltl$, model checking is still $\pspace$-complete and solving games is $\twoexp$-complete. Again, these problems were solved by an application of the alternating-color technique. One has to overcome some technicalities, but the general proof technique is the same as for $\prompt$. 

The decision problems for the parameterized logics mentioned above are boundedness problems, e.g., one asks for an upper bound on the waiting times between requests and responses in case of the formula~$\G(q \rightarrow \F_{\le x} p)$. Recently, more general boundedness problems in logics and automata received a lot of attention to obtain decidable quantitative extensions of monadic second-order logic and better synthesis algorithms. In general, boundedness problems are undecidable for automata with counters, but become decidable if the acceptance conditions can refer to boundedness properties of the counters, but the transition relation has no access to counter values. Recent advances include logics and automata with bounds~\cite{Bojanczyk04,BojanczykColcombet06}, satisfiability algorithms for
these logics~\cite{Bojanczyk11, Bojanczyk14, BojanczykTorunczyk12, Boom11}, and regular
cost-functions~\cite{Colcombet09}. However, these formalisms, while very expressive, are intractable and thus not suitable for verification and synthesis. 
Thus, less expressive formalisms were studied that appear more suitable for practical applications, e.g., finitary parity~\cite{ChatterjeeHenzingerHorn09}, parity with costs~\cite{FijalkowZimmermann14}, energy-parity~\cite{ChatterjeeDoyen10}, mean-payoff-parity~\cite{ChatterjeeHenzingerJurdzinski05}, consumption games~\cite{BCKN12}, and the use of weighted automata for specifying quantitative properties~\cite{BCHJ09, CernyChatterjeeHenzingerRadhakrishnaSingh11}. In particular, the parity condition with cost is defined in graphs whose edges are weighted by natural numbers (interpreted as costs) and requires the existence of a bound~$b$ such that almost every occurrence of an odd color is followed by an occurrence of a larger even color such that the cost between these positions is at most $b$. Although strictly stronger than the classical parity condition, solving parity games with costs is as hard as solving parity games~\cite{MogaveroMS13}.

\subparagraph*{Our contribution:} We investigate parameterized temporal logics in a weighted setting similar to the one of parity conditions with costs: our graphs are equipped with cost-functions that label the edges with natural numbers and parameterized operators are now evaluated with respect to cost instead of time, i.e., the parameters bound the accumulated cost instead of the elapsed time. Thus, the formula~$\G( q \rightarrow \F_{\le x}p)$ requires that every request~$q$ is answered with cost at most $\alpha(x)$. We show the following results about $\pltl$ with costs ($\pltlc$):

First, we refined the alternating-color technique to the cost-setting, which requires to tackle some technical problems induced by the fact that accumulated cost, unlike time, does not increase in every step, e.g., if an edge with cost zero is traversed. 

Second, we show that Kupferman et al.'s proofs based on the alternating-color technique can be adapted to the cost-setting as well. For model-checking, we again obtain $\pspace$-completeness while solving games is still $\twoexp$-complete.

Third, we consider $\pldl$ with costs ($\pldlc$), which is defined as expected. Again, the complexity does not increase: model checking is $\pspace$-complete while solving games is $\twoexp$-complete. 

Fourth, we generalize both logics to a setting with multiple cost-functions. Now, the parameterized temporal operators have another parameter~$i$ that determines the cost-function under which they are evaluated. Even these extensions do not increase complexity: model checking is again $\pspace$-complete while solving games is still $\twoexp$-complete. 

Fifth, we also investigate model checking and solving games as an optimization problem, which is a very natural view on the problems, i.e., we are interested in computing the optimal variable valuation such that a given system satisfies a given specification. For $\pltlc$ and $\pldlc$, we show that the model checking optimization problem can be solved in polynomial space while the optimization problem for infinite games can be solved in triply-exponential time. These results are similar to the ones obtained for $
\pltl$~\cite{Zimmermann13}. In particular, the exponential gap between the decision and the optimization variant of solving infinite games exists already for $\pltl$.

All proofs omitted due to space restrictions can be found in the full version~\cite{Zimmermann15}.

%% file: defs.tex
Let $\Var$ be an infinite set of variables and let $P$ be a set of atomic propositions. The formulae of $\pltlc$ are given by the grammar
\begin{equation*}\varphi \cceq p \mid \neg p \mid \varphi \wedge \varphi \mid \varphi \vee
\varphi
  \mid \X \varphi \mid \varphi \U \varphi \mid \varphi \R \varphi \mid
  \mathbf{F}_{ \le z } \varphi \mid \mathbf{G}_{ \le z} \varphi
,\end{equation*}
where $p \in P$ and $z \in \Var$. We use the derived operators $\ttrue \coloneq p
\vee \neg p$ and $\ffalse \coloneq p \wedge \neg p$ for some fixed $p \in P$, $\F
\varphi \coloneq \ttrue \U \varphi$, and $\G \varphi \coloneq \ffalse \R \varphi$. Furthermore, we use $p \rightarrow \varphi$ and $\neg p \rightarrow \varphi$ as shorthand for $\neg p \vee \varphi$ and $p \vee \varphi$, respectively. Additional derived operators are introduced on page~\pageref{page_derivedops}.

The set of subformulae of a $\pltlc$ formula~$\varphi$ is denoted by $\cl( \varphi )$
and we define the size of $\varphi$ to be the cardinality of $\cl(\varphi)$.
Furthermore, we define $\varF( \varphi ) = \{ z\in \Var \mid \F_{\le z} \psi \in
\cl( \varphi) \}$ to be the set of variables parameterizing eventually operators in
$\varphi$, $\varG( \varphi ) = \{ z\in \Var \mid \G_{\le z} \psi \in \cl( \varphi)  \} $
to be the set of variables parameterizing always operators in $\varphi$, and set
$\var( \varphi ) = \varF( \varphi ) \cup \varG( \varphi )$. 

$\pltlc$ is evaluated on so-called cost-traces (traces for short) of the form 
$w = w_0\, c_0\, w_1\, c_1\, w_2\, c_2\, \cdots \in \left( \pow{P} \cdot\ \nats \right)^{ \omega }$, which encode the evolution of the system in terms of the atomic propositions that hold true in each time instance, and the cost of changing the system state. The cost of the trace~$w$ is defined as $\cst(w) = \sum_{j \ge 0}c_j$, which might be infinite. A finite cost-trace is required to begin and end with an element of $\pow{P}$. The cost~$\cst(w)$ of a finite cost-trace~$w = w_0 c_0 w_1 c_1 \cdots c_{n-1} w_n$ is defined as $\cst(w) = \sum_{j=0}^{n-1}c_j$. 

Furthermore, we require the existence of a distinguished atomic proposition~$\inc$ such that all cost-traces satisfy $c_j >0 $ if and only if $\inc \in w_{j+1}$, i.e., $\inc$ indicates that the last step had non-zero cost. We use the proposition~$\inc$ to reason about costs: for example, we are able to express whether a trace has cost~zero or $\infty$. In the following, we will ensure that all our systems only allow traces that satisfy this assumption. 

Also, to evaluate formulas we need to  instantiate the variables parameterizing the temporal operators. To this end, we define a variable valuation to be a
mapping~$\alpha\colon \Var \rightarrow \nats$. 
Now, we can define the model
relation between a cost-trace~$w = w_0\, c_0\, w_1\, c_1\, w_2\, c_2\, \cdots $, a
position~$n$ of $w$, a variable valuation~$\alpha$, and a~$\pltlc$ formula as
follows:
\begin{itemize}
\item $(w,n,\alpha)\models p$ if and only if  $p\in w_n$,

\item $(w,n,\alpha)\models\neg p$ if and only if  $p\notin
w_n$,

\item $(w,n,\alpha)\models\varphi\wedge\psi$ if and only if
$(w,n,\alpha)\models\varphi$ and
$(w,n,\alpha)\models\psi$,

\item $(w,n,\alpha)\models\varphi\vee\psi$ if and only if
$(w,n,\alpha)\models\varphi$ or $(w,n,\alpha)\models\psi$,

\item $(w,n,\alpha)\models\X\varphi$ if and only if  $(w,n+1,\alpha)\models\varphi$,

\item $(w,n,\alpha)\models\varphi\U\psi$ if and only if there exists a $j\ge 0$
such that
$(w,n+j,\alpha)\models\psi$ and $(w,n+k,\alpha)\models\varphi$ for every $k$ in the
range $0\le k< j $,

\item $(w,n,\alpha)\models\varphi\R\psi$ if and only if  for every $j\ge 0$: either
$(w,n+j,\alpha)\models\psi$
or there exists a $k$ in the range $0\le k < j$ such that
$(w,n+k,\alpha)\models\varphi$,

\item $(w,n,\alpha)\models\F_{\le z}\varphi$ if and only if there exists a $j \ge 0$ with $\cst(w_{n} c_n \cdots c_{n+j-1} w_{n+j}) \le \alpha(z)$ such that $(w,n+j,\alpha)\models\varphi$, and

\item $(w,n,\alpha)\models\G_{\le z}\varphi$ if and only if for every $j \ge 0$ with $\cst(w_{n} c_n \cdots c_{n+j-1} w_{n+j}) \le \alpha(z)$: $(w,n+j,\alpha)\models\varphi$.

\end{itemize}
Note that we recover the semantics of $\pltl$ as the special case where every~$c_n$ is equal to one. 

For the sake of brevity, we write $(w,\alpha) \models \varphi$ instead of
$(w,0,\alpha) \models \varphi$ and say that $w$ is a model of $\varphi$ with
respect to $\alpha$. For variable-free formulas, we even drop the $\alpha$ and write $w \models \alpha$.

As usual for parameterized temporal logics, the use of variables has to be
restricted: bounding eventually and always operators by the same variable leads
to an undecidable satisfiability problem~\cite{AlurEtessamiLaTorrePeled01}.

\begin{definition}
\label{def_wellformedformula}
A $\pltlc$ formula~$\varphi$ is well-formed, if $\varF( \varphi ) \cap \varG( \varphi ) =
\emptyset$.
\end{definition}

In the following, we only consider well-formed formulas and omit the qualifier~\myquot{well-formed}. Also, we will denote variables in $\varF( \varphi )$ by $x$ and variables in $\varG( \varphi )$ by $y$, if the formula~$\varphi$ is clear from context.

We consider the following fragments of $\pltlc$. Let $\varphi$ be a $\pltlc$ formula:
\begin{itemize}
\item $\varphi$ is an $\ltl$ formula, if $\var( \varphi ) = \emptyset$.

\item $\varphi$ is a $\pltlcf$ formula, if $\varG(\varphi) = \emptyset$.

\item $\varphi$ is a $\pltlcg$ formula, if $\varF(\varphi) = \emptyset$.

\end{itemize}

\begin{example}\hfill
\begin{enumerate}
	\item The formula~$\G (q \rightarrow \F_{\le x}p)$ is satisfied with respect to $\alpha$, if every request (a position where $q$ holds) is followed by a response (a position where $p$ holds) such that the cost of the infix between the request and the response is at most $\alpha(x)$.
	\item The (max-) parity condition with costs~\cite{FijalkowZimmermann14} can be expressed in $\pltlc$ via
\[
\F\G  \left( \bigwedge\nolimits_{c \in \set{1, 3, \ldots, d-1}} \left(c \rightarrow  \F{}_{\le x}\bigvee\nolimits_{c' \in \set{c+1, c+3, \ldots, d}} c' 
\right)\right),
\]
where $d$ is the maximal color, which we assume w.l.o.g.\ to be even. However, the Streett condition with costs~\cite{FijalkowZimmermann14} cannot be expressed in $\pltlc$, as it is defined with respect to multiple cost functions, one for each Streett pair. We extend $\pltlc$ to multiple cost functions in Section~\ref{sec_mult}.

\end{enumerate}
\end{example}

As for $\pltl$, one can also parameterize the until and the release operator and also consider bounds of the form~\myquot{$> z$}. However, this does not increase expressiveness of the logic. Thus, we introduce these operators by defining them using $\F_{\le x}$ and $\G_{\le y}$:\label{page_derivedops}
\medskip

\begin{minipage}[b]{.4\textwidth}
	\begin{itemize}
	\item $\varphi \U_{\le x} \psi \coloneq \varphi \U \psi \wedge \F_{\le x} \psi$
\item 	$\varphi \R_{\le y} \psi \coloneq \varphi \R \psi \vee \G_{\le y} \psi$
\item 	$\F_{> y} \varphi \coloneq \G_{\le y}\F\X(\inc \wedge \F\varphi)$
	\end{itemize}
	\end{minipage}
\begin{minipage}[b]{.45\textwidth}
	\begin{itemize}
\item	$\G_{> x} \varphi \coloneq \F_{\le x}\G\X(\neg \inc \vee \G\varphi)$
\item	$\varphi \U_{> y} \psi \coloneq \G_{\le y}(\varphi \wedge \F\X(\inc \wedge \varphi\U\psi))$
\item $	\varphi \R_{> x} \psi \coloneq \F_{\le x}(\varphi \vee   \G\X(\neg \inc \vee \varphi\R\psi))$

	\end{itemize}
	\end{minipage}\medskip

Note that we defined $\pltlc$ formulae to be in negation normal form.
Nevertheless, a negation can be pushed to the atomic propositions using the
duality of the operators. Thus, we can define the negation of a $\pltlc$
formula.

\begin{lemma}
\label{lemma_pltlnegation}
For every $\pltlc$ formula~$\varphi$ there exists an efficiently constructible
$\pltlc$ formula~$\neg \varphi$ s.t.\
\begin{enumerate}
\item $(w,n,\alpha)\models \varphi$ if and only if $(w,n,
\alpha) \not\models \neg \varphi$ for every $w$, every $n$, and every $\alpha$,
\item $\card{\neg \varphi} =  \card{\varphi}$.
\item If $\varphi$ is well-formed, then so is $\neg \varphi$.
\item If $\varphi$ is an $\ltl$ formula, then so is $\neg \varphi$.
\item If $\varphi$ is a $\pltlcf$ formula, then $\neg \varphi$ is a $\pltlcg$ formula
and vice versa.
\end{enumerate}
\end{lemma}

%
%
%
%

Another important property of parameterized logics is monotonicity: increasing (decreasing) the values of parameterized eventuality operators (parameterized always operators) preserves satisfaction.

\begin{lemma}
\label{lemma_monotonicity}
Let $\varphi$ be a $\pltlc$ formula and let $\alpha$ and
$\beta$ be variable valuations satisfying $\beta ( x) \ge \alpha ( x )$ for
every $x \in \varF( \varphi)$ and $\beta ( y) \le \alpha ( y )$ for every $y \in \varG( \varphi)$. If $(w, \alpha) \models \varphi$, then $(w, \beta) \models
\varphi$.
\end{lemma}

Especially, if we are interested in checking whether a formula is satisfied with respect to some $\alpha$, we can always recursively replace every subformula~$\G_{\le y}\psi$ by 
$\psi \vee \X (\neg \inc \U (\neg \inc \wedge \psi))$, as this is equivalent to $\G_{\le y}\psi$ with respect to every variable valuation mapping $y$ to zero, which is the smallest possible value for $y$. Note that we have to ignore the current truth value of $\inc$, as it indicates the cost of the last transition, not the cost of the next one.

%% file: altcolor.tex
Fix a fresh atomic proposition~$p \notin P$. We say that a cost-trace~$w' = w_0' c_0' w_1' c_1' w_2' c_2' \cdots \in \left( \pow{P \cup \set{p}} \cdot \nats \right)^\omega$ is a coloring of a cost trace~$w = w_0 c_0 w_1 c_1 w_2 c_2 \cdots \in \left( \pow{P}  \cdot \nats \right)^\omega$, if $w_n' \cap P = w_n$ and $c_n' = c_n$ for every $n$, i.e., $w'$ and $w$ only differ in the truth values of the new proposition~$p$. A position~$n$ is a changepoint of $w'$, if $n=0$ or if the truth value of $p$ in $w_{n-1}'$ and $w_{n}'$ differs. A block of $w'$ is an infix $w_n' c_n' \cdots w_{n+j}'$ of $w'$ such that $n$ and $n+j+1$ are successive changepoints. If a coloring has only finitely many changepoints, then we refer to its suffix starting at the last changepoint as its tail, i.e., the coloring is the concatenation of a finite number of blocks and its tail. 

Let $k \in \nats$. We say that $w'$ is $k$-bounded if every block and its tail (if it has one) has cost at most $k$. Dually, we say that $w'$ is $k$-spaced, if every block has cost at least $k$. Note that we do not have a requirement on the cost of the tail in this case.

Given a $\pltlcf$ formula~$\varphi$, let $\rel{\varphi}$ denote the $\ltl$ formula obtained from $\varphi$ by recursively replacing every subformula~$\F_{\le x} \psi$ by
\[(p \rightarrow p\U(\neg p \U \rel{\psi})) \,\, \wedge \,\, 
(\neg p \rightarrow \neg p\U( p \U \rel{\psi})).
\]
Intuitively, the relativized formula requires $\rel{\psi}$ to be satisfied within at most one changepoint. On bounded and spaced colorings, $\varphi$ and $\rel{\varphi}$ are \myquot{equivalent}.

\begin{lemma}[cp.\ Lemma 2.1 of \cite{KupfermanPitermanVardi09}]
\label{lemma_altcolor}
Let $w$ be a cost-trace and let $\varphi$ be a $\pltlcf$ formula.
\begin{enumerate}
	
\item\label{lemma_altcolor_pltl2ltl}
Let $(w, \alpha) \models \varphi$ for some variable valuation $\alpha$. Then, $w' \models \rel{\varphi}$ for every $(k+1)$-spaced coloring $w'$ of $w$, where $k = \max_{x \in \var(\varphi)}\alpha(x)$.

\item\label{lemma_altcolor_ltl2pltl}
Let $w' \models \rel{\varphi}$ for some $k$-bounded coloring $w'$ of $w$. Then, $(w, \alpha ) \models \varphi$, where $\alpha(x) = 2k$ for every $x$.

\end{enumerate}
\end{lemma}

%% file: mc.tex
A transition system~$\sys = (S, s_I, E, \ell , \cst)$ consists of a finite directed graph~$(S, E)$, an initial state~$s_I \in S$, a labeling function~$\ell \colon S \rightarrow \pow{P}$, and a cost function\footnote{We encode the weights in binary, although the algorithms we present are oblivious to the exact values of the weights.}~$\cst \colon E \rightarrow \nats$. We assume that every state has at least one successor to spare us from dealing with finite paths. Recall our requirement on cost-traces having a distinguished atomic property~$\inc$ indicating the sign of the cost of the previous transition. Thus, we require $\sys$ to satisfy the following property: if $\inc \in \ell(v')$, then $\cst(v,v') >0$ for every edge~$(v,v') \in E$ leading to $v'$. Dually, if $\inc \notin \ell(v')$, then $\cst(v,v') = 0$ for every edge~$(v,v') \in E$.

A path through $\sys$ is a sequence~$\pi = s_0 s_1 s_2 \cdots$ satisfying $s_0 = s_I$ and $(s_n, s_{n+1}) \in E$ for  every $n$. Its cost-trace~$\trace(\pi)$ is defined as
\[
\trace(\pi) = 
\ell(s_0) \cst(s_0, s_1)
\ell(s_1) \cst(s_1, s_2)
\ell(s_2) \cst(s_2, s_3)
\cdots,
\]
which satisfies our assumption on the proposition~$\inc$.

The transition system~$\sys$ satisfies a $\pltlc$ formula~$\varphi$ with respect to a variable valuation~$\alpha$, if the trace of every path through $\sys$ satisfies $\varphi$ with respect to $\alpha$. The $\pltlc$ model checking problem asks, given a transition system~$\sys$ and a $\pltlc$ formula~$\varphi$, whether $\sys$ satisfies $\varphi$ with respect to some~$\alpha$.

\begin{theorem}
\label{thm_mc}
The $\pltlc$ model checking problem is $\pspace$-complete.
\end{theorem}

The proof we give below is a generalization of the one for $\prompt$ by Kupferman et al.~\cite{KupfermanPitermanVardi09}. We begin by showing  $\pspace$-membership. First note that we can restrict ourselves to $\pltlcf$ formulas: given a $\pltlc$ formula~$\varphi$, let $\varphi'$ denote the formula obtained by recursively replacing every subformula $\G_{\le y} \psi$ by $\psi \vee \X (\neg \inc \U (\neg \inc \wedge \psi))$. Due to Lemma~\ref{lemma_monotonicity} and the discussion below it, every transition system~$\sys$ satisfies $\varphi$ with respect to some $\alpha$ if and only if $\sys$ satisfies $\varphi'$ with respect to some~$\alpha'$.

Recall that $p$ is the distinguished atomic proposition used to relativize $\pltlc$ formulas. A colored Büchi graph with costs~$(V, v_I, E, \ell, \cst, F)$ consists of a finite direct graph~$(V, E)$, an initial vertex~$v_I$, a labeling function~$\ell \colon V \rightarrow \pow{\set{p}}$, a cost-function~$\cst \colon E \rightarrow \nats$, and a set~$F \subseteq V$ of accepting vertices. A path $v_0 v_1 v_2 \cdots $ is pumpable, if each of its blocks induced by $p$ contains a vertex repetition such that the cycle formed by the repetition has non-zero cost\footnote{Note that our definition is more involved than the one of Kupferman et al., since we require a cycle with non-zero cost instead of any circle.}. Note that we do not have a requirement on the cost of the tail, if the path has one. The path is fair, if it visits $F$ infinitely often. The pumpable non-emptiness problem asks, given a colored Büchi graph with costs, whether it has an initial pumpable fair path. 

\begin{lemma}
\label{lemma_upnonemptiness}
If a colored Büchi graph with costs has an initial pumpable fair path, then also one of the form~$\pi_0 \pi_1^\omega$ with $\size{\pi_0\pi_1} \in \mathcal{O}(n^2)$, where $n$ is the number of vertices of the graph.
\end{lemma}

Let $\sys = (S, s_I, E, \ell, \cst)$ be a transition system and let $
\varphi$ be a $\pltlcf$ formula. Furthermore, consider the $\ltl$ formula~$\chi = (\G\F p \wedge \G\F \neg p) \leftrightarrow \G\F \inc$,
which is satisfied by a cost-trace, if the trace has infinitely many changepoints if and only if it has cost~$\infty$. 
 Now, let $\aut = (Q, \pow{P \cup \set{p}}, q_I, \delta, F)$ be a nondeterministic Büchi automaton recognizing the models of the $\ltl$ formula~$\neg \rel{\varphi} \wedge \chi$, which we can pick such that its number of states is bounded exponentially in $\size{\varphi}$.
 Now, define the colored Büchi graph with costs~$\sys \times \aut = (S \times Q \times \pow{\set{p}}, (s_I, q_I, \emptyset), E', \ell', \cst', F')$ where
\begin{itemize}
	\item $((s,q, C),(s',q', C')) \in E'$ if and only if $(s,s') \in E$ and $q' \in \delta(q, \ell(s)\cup C)$,
	\item $\ell(s,q,C) = C$,
	\item $\cst'((s,q,C),(s',q',C')) = \cst(s,s') $, and
	\item $F' = S \times F \times \pow{\set{p}}$.
\end{itemize}

\begin{lemma}
\label{lemma_mccharacterization}[cp.\ Lemma 4.2 of \cite{KupfermanPitermanVardi09}]
$\sys$ does not satisfy $\varphi$ with respect to any $\alpha$ if and only if $\sys \times \aut$ has an initial pumpable fair path.
\end{lemma}

Now, we are ready to prove Theorem~\ref{thm_mc}.

\begin{proof}
$\pspace$-hardness holds already for $\ltl$~\cite{SistlaClarke85}, which is a fragment of $\pltlc$. Membership is witnessed by the following algorithm: check whether the colored Büchi graph~$\sys \times \aut$ has an initial pumpable fair path, which is correct due to Lemma~\ref{lemma_mccharacterization}. But as the graph is of exponential size, it has to be constructed and tested for non-emptiness on-the-fly.

Due to Lemma~\ref{lemma_upnonemptiness}, it suffices to check for the existence of an ultimately periodic path~$\pi_0\pi_1^\omega$ such that $\size{\pi_0\pi_1}\le n \in \mathcal{O}(\size{\sys \times \aut})$, i.e., $n$ is  exponential in the size of $\varphi$ and linear in the size of $\sys$. To this end, one guesses a vertex $v$ (the first vertex of $\pi_1$) and checks the following reachability properties:
\begin{enumerate}
	\item\label{item_pi1} Is $v$ reachable from $v_I$ via a path where each block contains a cycle with non-zero cost?
	\item Is $v$ reachable from $v$ via a non-empty path that visits an accepting vertex and which either has no changepoint or where each block contains a cycle with non-zero cost? In this case, we also require that $v$ and the last vertex on the path from $v_I$ to $v$ guessed in item \ref{item_pi1}.) differ on their third component in order to make $v$ a changepoint. This spares us from having a block that spans $\pi_0$ and $\pi_1$. 
\end{enumerate}
All these reachability problems can be solved in non-deterministic polynomial space, as a successor of a vertex of $\sys \times \aut$ can be guessed and verified in polymonial time and the length of the paths to be guessed is bounded by $n$, which can be represented with  polynomially many bits. 
\end{proof}

Furthermore, from the proof of Lemma~\ref{lemma_mccharacterization}, we obtain an exponential upper bound on the values of a satisfying variable valuation, if one exists. This is asymptotically tight, as one can already show exponential lower bounds for $\prompt$~\cite{KupfermanPitermanVardi09}.

\begin{corollary}
\label{cor_mcub}
Fix a transition system~$\sys$ and a $\pltlc$-formula~$\varphi$ such that $\sys$ satisfies $\varphi$ with respect to some $\alpha$. Then, $\sys$ satisfies $\varphi$ with respect to a valuation that is bounded exponentially in the size of $\varphi$ and linearly in the number of states of $\sys$ and in the maximal cost in $\sys$.
\end{corollary}

Dually, using pumping arguments one can show  the existence of an exponential variable valuation that witnesses whether a given $\pltlcg$ specification is satisfied with respect to every variable valuation.

\begin{lemma}
\label{lemma_ubmcalways}
Fix a transition system~$\sys$ and a $\pltlcg$-formula~$\varphi$ such that $\sys$ does not satisfy $\varphi$ with respect to every $\alpha$. Then, $\sys$ does not satisfy $\varphi$ with respect to a valuation that is bounded exponentially in the size of $\varphi$ and linearly in the number of states of $\sys$ and in the maximal cost in $\sys$.
\end{lemma}

The proof of the preceding Lemma is similar to the one of Lemma 7 in \cite{FaymonvilleZimmermann15}.

%% file: games.tex
An arena~$\arena = (V, V_0, V_1, v_I, E, \ell, \cst)$ consists of a finite directed graph~$(V, E)$, a partition~$(V_0, V_1)$ of $V$, an initial vertex~$v_I \in V$, a labeling~$\ell \colon V \rightarrow \pow{P}$, and a cost function\footnote{Again, we encode the weights in binary, although the algorithms we present are oblivious to the exact values of the weights.}~$\cst \colon E \rightarrow \nats$. Again, we assume that every vertex has at least one successor to avoid dealing with finite paths. Also, we again ensure our requirement on the proposition~$\inc$ to indicate the sign of the costs in a cost-trace: if $\inc \in \ell(v')$, then we require $\cst(v,v') >0$ for every edge~$(v,v') \in E$ leading to $v'$. Dually, if $\inc \notin \ell(v')$, then $\cst(v,v') = 0$ for every edge~$(v,v') \in E$.

A play~$\rho = \rho_0 \rho_1 \rho_2 \cdots$ is a path through $\arena$ starting in $v_I$ and its cost-trace~$\trace(\rho)$ is defined as
\[
\trace(\rho)  = 
\ell(\rho_0)\, \cst(\rho_0, \rho_1)\,
\ell(\rho_1)\, \cst(\rho_1, \rho_2)\,
\ell(\rho_2)\, \cst(\rho_2, \rho_3)
\cdots.\]

 A strategy for Player~$i \in \set{0,1}$ is a mapping~$\sigma \colon V^*V_i \rightarrow V$ satisfying $(v, \sigma(wv)) \in E$ for every $w \in V^*$ and $v \in V_i$. A play~$\rho$ is consistent with $\sigma$ if $\rho_{n+1} = \sigma (\rho_0 \cdots \rho_n)$ for every $n$ with $\rho_n \in V_i$.

A $\pltlc$ game~$\game = (\arena, \varphi)$ consists of an arena~$\arena$ and a winning condition~$\varphi$, which is a $\pltlc$ formula. A strategy~$\sigma$ for Player~$0$ is winning with respect to some variable valuation $\alpha$, if the trace of every play that is consistent with $\sigma$ satisfies the winning condition~$\varphi$ with respect to $\alpha$.

We are interested in determining whether Player~$0$ has a winning strategy for a given $\pltlc$~game, and in determining a winning strategy for her if this is the case. 

\begin{theorem}
\label{thm_games}
Determining whether Player~$0$ has a winning strategy in a given $\pltlc$ game is $\twoexp$-complete. Furthermore, a winning strategy (if one exists) can be computed in doubly-exponential time. 
\end{theorem}

Our proof technique is a generalization of the one for infinite games with $\pltl$ winning conditions~\cite{Zimmermann13}, which in turn extended Kupferman et al.'s solution for the $\prompt$ realizability problem~\cite{KupfermanPitermanVardi09}. First, we note that it is again sufficient to consider $\pltlcf$ formulas, as we are interested in the existence of a variable valuation (see the discussion below Lemma~\ref{lemma_monotonicity}). Next, we apply the alternating-color technique: to this end, we modify the arena to allow Player~$0$ to produce colorings of plays of the original arena and use the relativized winning condition, i.e., we reduce the problem to a game with $\ltl$ winning condition. The winner (and a winning strategy) of such a game can be computed in doubly-exponential time~\cite{PnueliRosner89,PnueliRosner89a}. 

To allow for the coloring, we double the vertices of the arena, additionally label one copy with $p$ and the other not, and split every move into two: first, the player whose turn it is picks an outgoing edge, then Player~$0$ decides in which copy she wants to visit the target, thereby picking the truth value of $p$.

Formally, given $\arena=(V,V_0,V_1, v_I, E,\ell, \cst)$, the extended arena $\arena' =
(V',V_0',V_1', v_I', E',\ell', \cst')$ consists of 
\begin{itemize}
\item $V'=V\times\{0,1\}\cup E$,
\item $V_0'=V_0\times\{0,1\}\cup E$ and $V_1'=V_1\times\{0,1\}$,
\item $v_I' = (v_I, 0)$,
\item $E'=\{((v,0),e),((v,1),e),(e,(v',0)),(e,(v',1))\mid e=(v,v')\in E\}$,
\item $\ell'(e)=\emptyset$ for every $e\in E$ and
$\ell'(v,b)=\begin{cases}\ell(v) & \text{if $b=0$},\\
{\ell(v)\cup\{p\}}&\text{if $b=1$},\\
         \end{cases}$ and
\item $\cst'((v,b),(v,v')) = \cst(v,v') $ and $\cst'((v,v'),(v',b'))=0$.
\end{itemize}

A path through $\arena'$ has the
form $(\rho_0,b_0)e_0(\rho_1,b_1)e_1(\rho_2,b_2)\cdots$ for some path
$\rho_0\rho_1\rho_2\cdots$ through $\arena$, where  $e_n=(\rho_n,\rho_{n+1})$ and $b_n \in \{0,1\}$. Also, we have
$\card{\arena'}\in\mathcal{O}(\card{\arena}^2)$. Note that we use the costs in $\arena'$ only to argue the correctness of our construction, not to define the winning condition for the game in $\arena'$. 
 
Also note that the additional choice vertices of the form $e \in E$ have to be ignored when it comes to evaluating the winning condition on the trace of a play. Thus, we consider games with $\ltl$ winning conditions under so-called \emph{blinking semantics}: Player~$0$ wins a play~$\rho = \rho_0 \rho_1 \rho_2 \cdots$ under blinking semantics, if $\ell(\rho_0)\ell(\rho_2)\ell(\rho_4) \cdots$ satisfies the winning condition~$\varphi$; otherwise, Player~$1$ wins. Winning strategies under blinking semantics are defined as expected. Determining whether Player~$0$ has a winning strategy for a given game with $\ltl$ winning condition under blinking semantics is $\twoexp$-complete, which can be shown by a slight variation of the proof for $\ltl$ games under classical semantics~\cite{PnueliRosner89,PnueliRosner89a}. Furthermore, if Player~$0$ has a winning strategy for such a game, then also a finite-state one of at most doubly-exponential size in $\size{\varphi}$. 

Such a strategy for an arena~$(V, V_0, V_1, v_I, E, \ell, \cst)$ is given by a memory structure~$\mem = (M, m_I, \update)$ with a finite set~$M$ of memory states, an initial memory state~$m_I \in M$, and an update function~$\update\colon M \times V \rightarrow M$, and by a next-move function~$\nextmove \colon V_0 \times M \rightarrow V$ satisfying $(v, \nextmove(v,m)) \in E$ for every $m$ and every $v$. The function~$\update^* \colon V^+ \rightarrow M$ is defined via $\update^*(v) = m_I$ and $\update^*(wv) = \update(\update^*(w),v)$. Then, the strategy~$\sigma$ implemented by $\mem$ and $\nextmove$ is defined by $\sigma(wv) = \nextmove(v, \update^*(wv))$. The size of $\sigma$ is (slightly abusively) defined as $\size{M}$.

Given a game~$(\arena, \varphi)$ with $\pltlcf$ winning condition~$\varphi$, define $\arena'$ as above and let $\varphi' = \rel{\varphi} \wedge \chi$, where 
$\chi = (\G\F p \wedge \G\F \neg p) \leftrightarrow \G\F \inc$. 
Recall that $\chi$ is satisfied by a cost-trace, if the trace has infinitely many changepoints if and only if it has cost~$\infty$.
 
\begin{lemma}
\label{lemma_gamescharacterization}[cp.\ Lemma 3.1 of \cite{KupfermanPitermanVardi09}]
Player~$0$ has a winning strategy for $(\arena, \varphi)$ with respect to some $\alpha$ if and only if she has a winning strategy for $(
\arena',\varphi')$ under blinking semantics.
\end{lemma}

Now, we are able to prove Theorem~\ref{thm_games}.

\begin{proof}
Hardness follows immediately from the $\twoexp$-hardness of determining the winner of an $\ltl$ game~\cite{PnueliRosner89,PnueliRosner89a}, as $\ltl$ is a fragment of $\pltlc$.

Membership in $\twoexp$ follows from the reductions described above: first, we turn the winning condition into a $\pltlcf$ formula and construct the $\ltl$ game under blinking semantics obtained from expanding the arena and relativizing the winning condition. This game is only polynomially larger than the original one and its winner (and a winning strategy) is computable in doubly-exponential~time. 
\end{proof}

From the proof of Lemma~\ref{lemma_gamescharacterization}, we obtain a doubly-exponential upper bound on the values of a satisfying variable valuation, if one exists. This is asymptotically tight, as one can already show doubly-exponential lower bounds for $\prompt$~\cite{Zimmermann13}.

\begin{corollary}
\label{cor_gamesub}
Fix a $\pltlc$ game~$\game = (\arena, \varphi)$ such that Player~$0$ has a winning strategy for $\game$ with respect to some $\alpha$. Then, Player~$0$ has a winning strategy for $\game$ with respect to a valuation that is bounded doubly-exponentially in the size of $\varphi$ and linearly in the number of vertices of $\arena$ and in the maximal cost in $\arena$.
\end{corollary}

%% file: cpldl.tex
Linear Dynamic logic ($\ldl$)~\cite{GiacomoVardi13,Vardi11} extends $\ltl$ by temporal operators guarded with regular expressions, e.g., $\ddiamond{r}\varphi$ holds at position~$n$, if there is a $j$ such that $\varphi$ holds at position~$n+j$ and the infix between positions~$n$ and $n+j$ matches $r$. The resulting logic has the full expressiveness of the $\omega$-regular languages while retaining many of $\ltl$'s desirable properties like a simple syntax, intuitive semantics, a polynomial space algorithm for model checking, and a doubly-exponential time algorithm for solving games. Parametric $\ldl$ ($\pldl$)~\cite{FaymonvilleZimmermann14} allows to parameterize such operators, i.e., $\ddiamondle{r}{x}\varphi$ holds at position~$n$ with respect to a variable valuation~$\alpha$, if there is a $j \le \alpha(x)$ such that $\varphi$ holds at position~$n+j$ and the infix between positions~$n$ and $n+j$ matches $r$. Model checking and solving games with $\pldl$ specifications is not harder than for $\ltl$, although $\pldl$ is more expressive and has parameterized operators. In this section, we consider $\pldlc$ where the parameters bound the cost of the infix instead of the length. 

Formally, formulas of $\pldlc$ are given by the grammar
\begin{align*}
\varphi &\cceq p \mid \neg p \mid \varphi \wedge \varphi \mid \varphi \vee \varphi
  \mid \ddiamond{r} \varphi 
  \mid \bbox{r} \varphi 
  \mid \ddiamondle{r}{z} \varphi 
  \mid \bboxle{r}{z} \varphi\\
  r & \cceq \phi \mid \varphi? \mid r+r \mid r \conc r \mid r^*
\end{align*}
where $p \in P$, $z \in \Var$, and where $\phi$ ranges over propositional formulas over $P$. As for $\pltlc$, $\pldlc$ formulas are evaluated on cost-traces with respect to variable valuations. Satisfaction of atomic formulas and of conjunctions and disjunctions is defined as usual, and for the four temporal operators, we define
\begin{itemize}

\item $(w, n, \alpha) \models \ddiamond{r}\varphi$ if there exists $j \ge 0$ such that $(n, n+j) \in \Rexp(r, w, \alpha)$ and $(w, n+j, \alpha) \models \varphi$, 

\item $(w, n, \alpha) \models \bbox{r}\varphi$ if for all $j \ge 0$ with $(n, n+j) \in \Rexp(r, w, \alpha)$ we have $(w, n+j, \alpha) \models \varphi$,

\item $(w, n, \alpha) \models \ddiamondle{r}{z}\varphi$ if there exists $j \ge 0$ with $\cst(w_n c_n \cdots c_{n+j-1}w_{n+j}) \le \alpha(z)$ such that $(n, n+j) \in \Rexp(r, w, \alpha)$ and $(w, n+j, \alpha) \models \varphi$, and

\item $(w, n, \alpha) \models \bboxle{r}{z}\varphi$ if for all $j \ge 0$ with $\cst(w_n c_n \cdots c_{n+j-1}w_{n+j}) \le \alpha(z)$ and with $(n, n+j) \in \Rexp(r, w, \alpha)$ we have $(w, n+j, \alpha) \models \varphi$.

\end{itemize}
Here, the relation~$\Rexp(r,w,\alpha) \subseteq \nats\times\nats$ contains all pairs~$(m,n)$ such that $w_m \cdots w_{n-1}$ matches $r$ and is defined inductively by 
\begin{itemize}
\item $\Rexp(\phi,w,\alpha) = \set{(n, n+1) \mid w_n \models \phi}$ for propositional~$\varphi$,
\item $\Rexp(\psi?,w,\alpha) = \set{(n, n) \mid (w, n, \alpha) \models \psi}$,
\item $\Rexp(r_0 + r_1, w, \alpha) = \Rexp(r_0, w, \alpha) \cup \Rexp(r_1, w, \alpha)$,
\item $\Rexp(r_0 \conc r_1, w, \alpha) = \set{(n_0, n_2) \mid \exists n_1 \text{ s.t. }(n_0,n_1)\in \Rexp(r_0, w, \alpha) \text{ and } (n_1, n_2) \in \Rexp(r_1, w, \alpha)}$, and 
\item $\Rexp(r^*, w, \alpha) = \set{(n,n) \mid n\in\nats} \cup \set{(n_0, n_{k+1}) \mid \exists n_1, \ldots, n_{k} \text{ s.t. } (n_j, n_{j+1}) \in \Rexp(r, w, \alpha) \text{ for all } j \le k}$.
\end{itemize}
Again, we restrict ourselves to formulas where the set of variables parameterizing diamond operators and the set of variables parameterizing box operators are disjoint. Analogues of Lemma~\ref{lemma_pltlnegation} and Lemma~\ref{lemma_monotonicity} hold for $\pldlc$, too. 

The alternating-color is applicable to $\pldl$~\cite{FaymonvilleZimmermann14}: to this end, one introduces changepoint-bounded variants of the diamond- and the box-operator whose semantics only quantify over infixes with at most one changepoint. $\ldl$ formulas with  changepoint-bounded operators can be translated into Büchi automata of exponential size. This allows to extend the algorithms for model-checking and realizability based on the alternating-color technique~\cite{KupfermanPitermanVardi09} to $\pldl$. Even more so, the algorithms presented in Section~\ref{sec_mc} and Section~\ref{sec_games} can easily be adapted to $\pldlc$ as well, again relying on the translation to Büchi automata via changepoint-bounded operators.

\begin{theorem}
The $\pldlc$ model checking problem is $\pspace$-complete and determining the winner of games with $\pldlc$ winning conditions is $\twoexp$-complete.
\end{theorem}

%% file: multcost.tex
In this section, we consider parameterized temporal logics with  multiple cost-functions. For the sake of simplicity, we restrict our attention to $\pltlc$, although all results hold for $\pldlc$, too.

Fix some dimension~$d \in \nats$. The syntax of $\mpltlc$ is obtained by equipping the parameterized temporal operators by a coordinate~$i \in \set{1, \ldots, d}$, denoted by $\F_{\le_i x}$ and $\G_{\le_i y}$. In this context, a cost-trace is of the form
$
w_0\, \overline{c}_0\, 
w_1\, \overline{c}_1\, 
w_2\, \overline{c}_2\, 
\cdots$
where $w_n \in \pow{P}$ and $\overline{c}_n \in \nats^d$. Thus, for every $i \in \set{1, \ldots, d}$, we can define~$\cst_i(w_0 \overline{c}_0 \cdots \overline{c}_{n-1} w_{n}) = \sum_{j=0}^{n-1} (\overline{c}_j)_i$ for every finite cost-trace~$w_0 \overline{c}_0 \cdots \overline{c}_{n-1} w_{n}$. Furthermore, we require for every coordinate~$i$ a proposition~$\inc_i$ such that $\inc_i \in w_{n+1}$ if and only if $(\overline{c}_n)_i >0$. 

The semantics of atomic formulas, boolean connectives, and unparameterized temporal operators are unchanged and for the parameterized operators, we define 
\begin{itemize}
	\item $(w,n,\alpha)\models\F_{\le_i z}\varphi$ if and only if there exists a $j \ge 0$ with $\cst_i(w_{n} \overline{c}_n \cdots \overline{c}_{n+j-1} w_{n+j}) \le \alpha(z)$ such that $(w,n+j,\alpha)\models\varphi$, and

\item $(w,n,\alpha)\models\G_{\le_i z}\varphi$ if and only if for every $j \ge 0$ with $\cst_i(w_{n} \overline{c}_n \cdots \overline{c}_{n+j-1} w_{n+j}) \le \alpha(z)$: $(w,n+j,\alpha)\models\varphi$.

\end{itemize}

Again, we restrict ourselves to formulas where no variable parameterizes an eventually- and an always-operator, but we allow a variable to parameterize operators with different coordinates. Analogues of Lemma~\ref{lemma_pltlnegation} and Lemma~\ref{lemma_monotonicity} hold for $\mpltlc$ as well. 

\begin{example}
A Streett condition with costs~$(Q_i, P_i)_{i \in \{1, \ldots, d\}}$~\cite{FijalkowZimmermann14} can be expressed in $\mpltlc$ via
\[
\F\G  \left( \bigwedge\nolimits_{i \in \set{1, \ldots, d}} \left(Q_i \rightarrow  \F{}_{\le_i x}\, P_i 
\right)\right).
\]

\end{example}

In this setting, we consider the model checking problem for transition systems with $d$ cost functions and want to solve games in arenas with $d$ cost functions.

The alternating-color technique is straightforwardly extendable to $\mpltlc$: one introduces a fresh proposition~$p_i$ for each coordinate~$i$ and defines~$\chi = \bigwedge_{i=1}^d ( (\G\F p_i \wedge \G\F \neg p_i) \leftrightarrow \G\F \inc_i )$. Furthermore, the notions of $i$-blocks, $k$-boundedness in coordinate~$i$, and $k$-spacedness in coordinate~$i$ are defined as expected. Then, the proofs presented in Section~\ref{sec_mc} and Section~\ref{sec_games} remain valid in this setting, too.

 In the case of model-checking, the third component of the colored Büchi graph~$\sys \times \aut$ has the form~$\pow{\set{p_1, \ldots, p_d}}$, i.e., it is exponential. However, this is no problem, as the automaton~$\aut$ is already of exponential size. Similarly, in the case of infinite games, each vertex of the original arena has $2^d$ copies in $\arena'$, one for each element in $\pow{\set{p_1, \ldots, p_d}}$ allowing Player~$0$ to produce appropriate colorings with the propositions~$p_i$. The resulting game has an arena of exponential size (in the size of the original arena and of the original winning condition) and an $\ltl$ winning condition under blinking semantics. Such a game can still be solved in doubly-exponential time. To this end, one turns the winning condition into a deterministic parity automaton of doubly-exponential size with exponentially many colors, constructs the product of the arena and the parity automaton, which yields a parity game of doubly-exponential size with exponentially many colors. Such a game can be solved in doubly-exponential time~\cite{Schewe07}. 

\begin{theorem}
The $\mpltlc$ model checking problem is $\pspace$-complete and determining the winner of games with $\mpltlc$ winning conditions is $\twoexp$-complete.
\end{theorem}

Again, the same results hold for $\mpldlc$, which is defined as expected. 

%% file: optimization.tex
It is natural to treat model checking and solving games with specifications in parameterized linear temporal logics as an optimization problem: determine the \emph{optimal} variable valuation such that the system satisfies the specification with respect to it. For parameterized eventualities, we are interested in minimizing the waiting times while for parameterized always', we are interested in maximizing the waiting times. Due to the undecidability results for not well-defined formulas one considers the optimization problems for the unipolar fragments, i.e., for formulas having either no parameterized eventualities or no parameterized always'. In this section, we present algorithms for such optimization problems given by $\pltlc$ specifications. In the following, we encode the weights of the transition system or arena under consideration in unary to obtain our results. Whether these results can also be shown for a binary encoding is an open question. 

For model checking, we are interested in the following four problems: given a transition system~$\sys$ and a $\pltlcf$ formula~$\varphi_{\F}$ and a $\pltlcg$ formula~$\varphi_{\G}$, respectively, determine
\begin{enumerate}
\item\label{item_minminmc} $\min_{\{\alpha\mid \text{$\sys$ satisfies $\varphi_{\F}$ w.r.t.\ $\alpha$}\}}\min_{x\in \varF(\varphi_{\F})}\alpha(x)$,

\item $\min_{\{\alpha\mid \text{$\sys$ satisfies $\varphi_{\F}$ w.r.t.\ $\alpha$}\}}\max_{x\in \varF(\varphi_{\F})}\alpha(x)$,

\item $\max_{\{\alpha\mid \text{$\sys$ satisfies $\varphi_{\G}$ w.r.t.\ $\alpha$}\}}\max_{y\in \varG(\varphi_{\G})}\alpha(y)$, and

\item $\max_{\{\alpha\mid \text{$\sys$ satisfies $\varphi_{\G}$ w.r.t.\ $\alpha$}\}}\min_{y\in \varG(\varphi_{\G})}\alpha(y)$.

\end{enumerate}

Applying the monotonicity of the parameterized operators and (in the first case) the alternating-color technique to all but one variable reduces the four optimization problems to ones where the specification has a single variable. Furthermore, the upper bounds presented in Corollary~\ref{cor_mcub} and in Lemma~\ref{lemma_ubmcalways} yield an exponential search space for an optimal valuation: if this space is empty, then there is no $\alpha$ such that $\sys$ satisfies $\varphi_{\F}$ with respect to $\alpha$ in the first two cases. On the other hand, if the search space contains every such $\alpha$, then $\sys$ satisfies $\varphi_{\G}$ with respect to every $\alpha$ in the latter two cases. 

Thus, it remains the check whether the specification is satisfied with respect to some valuation that is bounded exponentially. In this setting, one can construct an exponentially sized non-deterministic Büchi automaton recognizing the models of the specification with respect to the given valuation (using a slight adaption of the construction presented in~\cite{Zimmermann13} accounting for the fact that we keep track of cost instead of time). This automaton can be checked for non-emptiness in polynomial space using an on-the-fly construction. Thus, an optimal $\alpha$ can be found in polynomial space by binary search.

\begin{theorem}
The $\pltlc$ model checking optimization problems can be solved in polynomial space.
\end{theorem}

A similar approach works for infinite games as well. Here, we are interested in computing 
\begin{enumerate}
\item\label{item_minmingame} $\min_{\{\alpha\mid \text{Pl.
 $0$ has winning strategy for $\game_{\F}$ w.r.t.\ $\alpha$}\}}\min_{x\in \varF(\varphi_{\F})}\alpha(x)$,

\item $\min_{\{\alpha\mid \text{Pl.
 $0$ has winning strategy for $\game_{\F}$ w.r.t.\ $\alpha$}\}}\max_{x\in \varF(\varphi_{\F})}\alpha(x)$,

\item $\min_{\{\alpha\mid \text{Pl.
 $0$ has winning strategy for $\game_{\G}$ w.r.t.\ $\alpha$}\}}\min_{x\in \varG(\varphi_{\G})}\alpha(x)$, and

\item $\min_{\{\alpha\mid \text{Pl.
 $0$ has winning strategy for $\game_{\G}$ w.r.t.\ $\alpha$}\}}\max_{x\in \varG(\varphi_{\G})}\alpha(x)$.

\end{enumerate}
and witnessing winning strategies for given $\pltlc$ games $\game_{\F}$ with $\pltlcf$ winning condition $\varphi_{\F}$ and $\game_{\G}$ with $\pltlcg$ winning condition $\varphi_{\G}$.

Again, one can reduce these problems to  the case of winning conditions with a single variable and by applying determinacy of games with respect to a fixed valuation, it even suffices to consider the case of $\pltlcf$ winning conditions with a single variable, due to duality of games: swapping the players in a game with $\pltlcg$ winning condition yields a game with $\pltlcf$ winning condition. Corollary~\ref{cor_gamesub} gives a doubly-exponential upper bound on an optimal variable valuation. Hence, one can construct a deterministic parity automaton of  triply-exponential size with exponentially many colors recognizing the models of the specification with respect to a fixed variable valuation~$\alpha$ that is below the upper bound (again, see~\cite{Zimmermann13} for the construction). Player~$0$ wins the parity game played in the original arena but using the language of the automaton as winning condition if and only if she has a winning strategy for the $\pltlcf$ game with respect to $\alpha$. Such a parity game can be solved in triply-exponential time~\cite{Schewe07}.

\begin{theorem}
The $\pltlc$ optimization problems for infinite games are solvable in triply-exponential time.
\end{theorem}

Furthermore, the same results hold for $\pldlc$ using appropriate adaptions of the automata constructions presented in~\cite{FaymonvilleZimmermann14,FaymonvilleZimmermann15}. 

\begin{theorem}
The $\pldlc$	  model checking optimization problems can be solved in polynomial space and the $\pldlc$ optimization problems for infinite games can be solved in triply-exponential time. 
\end{theorem}

However, for parameterized logics with multiple cost-functions, these results do not remain valid, as one cannot reduce the optimization problems to ones with a single variable, as a variable may bound operators in different dimensions. Thus, one has to keep track multiple costs, which incurs an exponential blow-up when done naively. Whether this can be improved is an open question. 

%% file: conc.tex
We introduced parameterized temporal logics whose operators bound the accumulated cost instead of time as usual: $\pltlc$ and $\pldlc$ as well as their variants $\mpltlc$ and $\mpldlc$ with multiple cost functions retain the attractive algorithmic properties of $\ltl$ like a polynomial space model checking algorithm and a doubly-exponential time algorithm for solving infinite games. Even the optimization variants of these problems are not harder for $\pltlc$ and $\pldlc$ than for $\pltl$: polynomial space for model checking and triply-exponential time for solving games. However, it is open whether these problems are harder for logics with multiple cost functions. Another open question concerns the complexity of the optimization problem for infinite games: can these problems be solved in doubly-exponential time, i.e., is finding optimal variable valuations as hard as solving games? Note that this question is already open for $\pltl$. Finally, one could consider weights from some arbitrary semiring and corresponding weighted parameterized temporal logics. 